\documentclass[a4paper, conference]{IEEEtran}

\IEEEoverridecommandlockouts                              

\usepackage[utf8]{inputenc}

\usepackage{graphicx}
\usepackage{caption}
\usepackage{subcaption}
\usepackage{color,soul}
\usepackage{url}

\usepackage{xcolor,listings}
\usepackage{textcomp}
\usepackage[normalem]{ulem}

\usepackage{tabularx}

\title{\LARGE \bf
From Simple Features to Moving Features and Beyond?
}

\author{\IEEEauthorblockN{Anita Graser}
\IEEEauthorblockA{\textit{AIT Austrian Institute of Technology} \\
\textit{University of Salzburg }\\
Vienna / Salzburg, Austria \\
ORCID: 0000-0001-5361-2885}
\and
\IEEEauthorblockN{Esteban Zimányi}
\IEEEauthorblockA{\textit{Université libre de Bruxelles} \\
Brussels, Belgium \\
ORCID: 0000-0003-1843-5099}
\and
\IEEEauthorblockN{Krishna Chaitanya Bommakanti}
\IEEEauthorblockA{\textit{Adonmo}\\ 
HITEC City, Hyderabad, \\ Telangana 500081, India }
}

\begin{document}

\maketitle

\thispagestyle{empty}
\pagestyle{empty}

\begin{abstract}

Mobility data science lacks common data structures and analytical functions.
This position paper assesses the current status and open issues towards a universal API for mobility data science. In particular, we look at standardization efforts revolving around the OGC Moving Features standard which, so far, has not attracted much attention within the mobility data science community. We discuss the hurdles any universal API for movement data has to overcome and propose key steps of a roadmap that would provide the foundation for the development of this API. 

\end{abstract}

\section{Introduction}

Data analysis tools are essential for data science. Robust movement data analysis tools are therefore key to advancing mobility data science. However, the development of movement data analysis tools is hampered by a lack of shared understanding and standardization. There are numerous implementations, including dozens of R libraries, as well as Python libraries and moving object databases. For example, \cite{joo2020navigating} review 57 R libraries related to movement in ecology. The development of Python libraries for movement analysis is picking up as well. For example, \cite{graser_movingpandas_2019} introduces the Python library MovingPandas and compares it to the R trajectories library \cite{moradi2018trajectories} and the movement database Hermes \cite{pelekis2015hermes}. Other Python libraries for movement analysis include scikit-mobility (focusing on human movement) \cite{pappalardo2019scikitmobility} and Traja (for animal movement) \cite{justin_shenk_2019_3237827}. However, even though they are all built for movement data analysis, they still vary considerably in their underlying concepts and provided functionality.

The ISO Standard 19141 Geographic information -- Schema for moving features was first published in 2008 \cite{ISO19141}. It defines a data model for representing moving points and moving rigid regions. Based on this standard, the Open Geospatial Consortium (OGC) has worked on various encodings for representing moving features as well as standardized operations for manipulating moving geometries. However, so far, these standards have failed to reach significant adoption. 

This discussion paper summarizes the current status as well as open issues towards a universal API for mobility data science. We start with a short introduction to OGC Moving Features standard and provide conceptual and technical commentary on these standardization efforts. These findings should serve as a point of departure for a community-wide effort to develop a shared understanding of what would be the right API for mobility data science -– if such a thing exists. Based on these findings, we then propose essential building blocks to advance movement data science.

\begin{figure}[tb]
\includegraphics[width=\columnwidth]{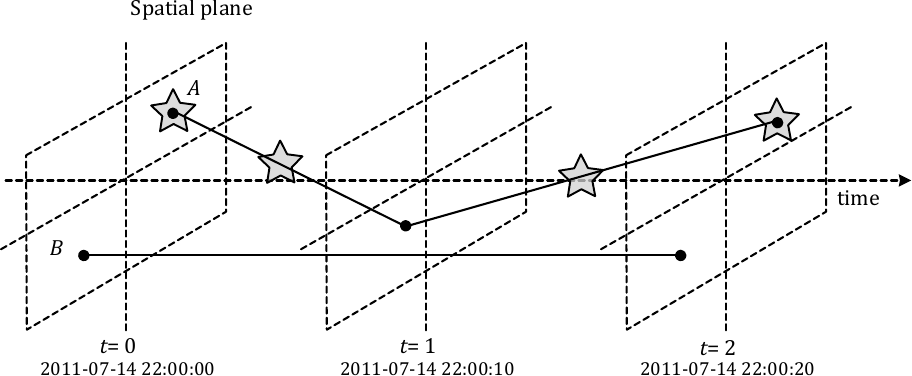}
\caption{Data model of the Moving Features standard illustrated with two moving points A and B. Stars mark changes in attribute values.}
\label{fig:example}
\end{figure}

\section{Moving Features}

The ISO and OGC Simple Features standard \cite{ISO19125} has gained widespread adoption in Geographic Information Systems (GIS) and related systems dealing with spatial data. Following this example, the new set of OGC Moving Features standards (which are based on the ISO standard \cite{ISO19141}) define how moving features should be encoded and how they should be accessed. A moving feature contains a temporal geometry, whose location changes over time, as well as dynamic non-spatial attributes whose values vary with time. The standards supports 0-dimensional (points), 1-dimensional (lines), 2-dimensional (polygons), and 3-dimensional (polyhedrons) geometries that vary over time. The standard allows the representation of the following phenomena: 

\begin{enumerate}
\item Discrete phenomena, which exist only on a set of instants, such as road accidents,
\item Step phenomena, where the changes of locations are abrupt at an instant, such as the location of mobile speed cameras for monitoring traffic, and
\item Continuous phenomena, whose locations move continuously for a period in time, such as vehicles, typhoons, or floods.
\end{enumerate}

\begin{table*}[h]
  \centering
  \begin{tabularx}{\textwidth}{lXXX}
    \hline
    & \multicolumn{1}{c}{\textbf{CSV}} 
    & \multicolumn{1}{c}{\textbf{XML}} 
    & \multicolumn{1}{c}{\textbf{JSON}} \\
    \hline
    \textbf{Concept}       
      & Segments with start and end time and attributes, no static object properties
      & Segments with start and end time and attributes, with static properties
      & Points with timestamps and attributes with (independent) timestamps 
      \\
    \hline
    \textbf{Advantages} 
      & \begin{enumerate}
            \item Supports temporal gaps in observations
        \end{enumerate}  
      & \begin{enumerate}
            \item Supports temporal gaps in observations
            \item Can handle complex geometries 
            \item Support for static / non-temporal descriptive object properties
        \end{enumerate} 
      & \begin{enumerate}
            \item Most compact representation
            \item Can handle complex geometries 
            \item Time stamps of location and attribute changes are modelled independently -- no synchronization necessary
            \item Interpolation modes can be specified individually for each attribute
            \item Support for non-temporal descriptive attributes
        \end{enumerate} 
        \\
    \hline
    \textbf{Limitations} 
      & \begin{enumerate}
            \item Verbose: redundant information (start and end time and location for each segment)
            \item Temporal geometry and temporal attributes have to be synced
            \item Only linear interpolation
            \item Only moving points
            \item Not readily usable in GIS (missed the opportunity to use WKT to represent the geometry)
        \end{enumerate}  
      & \begin{enumerate}
            \item Very verbose: XML \& redundant information (start and end time and location for each segment)
            \item Temporal geometry and temporal attributes have to be synced
            \item Same interpolation for geometry and all attributes
        \end{enumerate} 
      & \begin{enumerate}
            \item Multiple options for encoding the same situation (e.g. unclear bounds of time periods)
            \item No support for temporal gaps in observations
        \end{enumerate} 
        \\
    \hline
  \end{tabularx}
  \caption{Overview of OGC Moving Feature encodings}
  \label{tab:1}
\end{table*}

Fig.~\ref{fig:example} illustrates the Moving Features data model. It shows the movement of two vehicles denoted by $A$ and $B$. The goal is to model the vehicles' changing location and gear settings. The vehicle locations are modelled as moving points while the selected gear is modelled as a temporal property. Both vehicles start their movement at time $t=0$. While vehicle $A$ records its location at time $t=1$ and $t=2$, vehicle $B$ records its location only at time $t=2$. Furthermore, timestamps when $A$ changes gears are marked by a star symbol.

This example follows the default assumption of linear interpolation between locations. 
In general, a moving feature can be modelled as positions recorded at discrete timestamps, where the position between two recorded timestamps is computed by interpolation. If the interpolation is discrete, then no assumption is made about the location of the object between two observations. On the other hand, stepwise interpolation assumes that the location of the object is constant between two observations. Finally, linear or polynomial interpolations specify how to compute the location of the object between two observations. 

Based on this data model, OGC Simple Features defines XML/GML \cite{OGC14-083r2} and JSON \cite{OGC16-140r1} data encodings, as well as simpler CSV \cite{OGC14-084r2} and binary encodings (which are limited to moving points). However, even though these encodings are part of the same standard and are based on the same general data model, there are still considerable differences that significantly affect which kind of information can and can not be modelled (as shown in Table~\ref{tab:1}). To dive deeper and illustrate our point, the following sections show the example introduced in  Fig.~\ref{fig:example} encoded using the CSV, XML, and JSON encodings.

\subsection{CSV Encoding}

The CSV encoding is the simplest option in the Moving Features standard. 
Fig.~\ref{fig:csv} shows the CSV representation of the example introduced in Fig.~\ref{fig:example}. The CSV structure is divided into two parts: First, the header lines (starting with the ``\verb+@+" character) provide the meta-information. Then the trajectory lines describe the movement. The CSV encoding is segment based \cite{OGC14-084r2}, that is, each line describes a trajectory segment with two or more coordinate pairs (or triplets for 3D trajectories).

\begin{figure}[htb]
\begin{lstlisting}
@stboundedby,urn:x-ogc:def:crs:EPSG:6.6:4326,
	2D,10.0 10.0,10.6 12.2,2011-07-14T22:00:00Z,
	2011-07-14T22:00:20Z,sec
@columns,mfidref,trajectory,gear,xsd:integer
A,0,5,10.0 10.0 10.2 10.6,1
A,5,10,10.2 10.6 10.4 11.2,2
A,10,15,10.4 11.2 10.5 11.7,2
A,15,20,10.5 11.7 10.6 12.2,3
B,0,20,2.0 2.0 2.1 2.1,1
\end{lstlisting}
\caption{CSV encoding of vehicles $A$ and $B$}
\label{fig:csv}
\end{figure}

\begin{figure*}[htb]
\begin{lstlisting}[language=XML]
<?xml version="1.0" encoding="UTF-8"?>
	<mf:MovingFeatures xmlns:mf="http://schemas.opengis.net/mf-core/1.0" ...>
	<mf:STBoundedBy offset="sec">
		<gml:EnvelopeWithTimePeriod srsName="urn:x-ogc:def:crs:EPSG:6.6:4326">
			<gml:lowerCorner>10.0 10.0</gml:lowerCorner>
			<gml:upperCorner>10.6, 12.2</gml:upperCorner>
			<gml:beginPosition>2011-07-14T22:00:00Z</gml:beginPosition>
			<gml:endPosition>2011-07-14T22:00:20Z</gml:endPosition>
		</gml:EnvelopeWithTimePeriod>
	</mf:STBoundedBy>
	<mf:Member>
		<mf:MovingFeature gml:id="A">
			<gml:name>NissanA</gml:name>
			<gml:description>Nissan Sentra ...</gml:description>
		</mf:MovingFeature>
	</mf:Member>
	<mf:Header>
		<mf:VaryingAttrDefs>
			<mf:attrDef name="gear" type="xsd:integer">
			<mf:AttrAnnotation>The gear number used... </mf:AttrAnnotation>
			</mf:attrDef>
		</mf:VaryingAttrDefs>
	</mf:Header>
	<mf:Foliation>
		<mf:LinearTrajectory gml:id="LT0001" mfIdRef="A" start="0" end="5">
			<gml:posList>10.0 10.0 10.2 10.6</gml:posList>
			<mf:Attr>1</mf:Attr>
		</mf:LinearTrajectory>
		<mf:LinearTrajectory gml:id="LT0003" mfIdRef="A" start="5" end="10">
			<gml:posList>10.2 10.6 10.4 11.2</gml:posList>
			<mf:Attr>2</mf:Attr>
		</mf:LinearTrajectory>
		<mf:LinearTrajectory gml:id="LT0003" mfIdRef="A" start="10" end="15">
			<gml:posList>10.4 11.2 10.5 11.7</gml:posList>
			<mf:Attr>2</mf:Attr>
		</mf:LinearTrajectory>
		<mf:LinearTrajectory gml:id="LT0003" mfIdRef="A" start="15" end="20">
			<gml:posList>10.5 11.7 10.6 12.2</gml:posList>
			<mf:Attr>3</mf:Attr>
		</mf:LinearTrajectory>
	</mf:Foliation>
</mf:MovingFeatures>
\end{lstlisting}
\caption{XML encoding of vehicle $A$}
\label{fig:xml}
\end{figure*}

The first header line specifies the spatio-temporal boundary of the moving features. It lists the spatial reference system used, the number of dimensions of the geometries, the coordinates of the corners of the spatial boundary, the start and end times, and the time encoding, that is, the units of time used in the trajectory lines for encoding the offset from the start time. 

The second header line specifies the columns in the trajectory lines. The first column \verb+mfidref+ is used to identify the moving object. The second column \verb+trajectory+ defines the spatio-temporal geometry, followed by the definition of the time-varying attributes (name and its type). 

The following trajectory lines contain the spatio-temporal geometries for moving features. Each line specifies the \verb+mfidref+, the start and end time, which can be specified as absolute or relative (offset) values, the points of the line string and the attribute values. 
Since temporal geometry and temporal attributes (in this case the gear) are specified together, it is necessary to encode intermediate locations each time a temporal attribute changes its value. 

The CSV encoding assumes linear interpolation between locations. There is no way to store timestamps for intermediate positions along trajectory segments. The interpolation assumes constant speed along a segment. 

A notable design choice is that that the segment geometries are not encoded as well-known text (WKT), which would have made the CSV more easily readable by existing GIS tools.

\subsection{XML Encoding}

Like the CSV encoding, the XML encoding is also segment based. 
Fig.~\ref{fig:xml} shows the XML representation of the example introduced in Fig.~\ref{fig:example}. It is obvious that this encoding is considerably more verbose than the CSV encoding. The most essential XML tags are:

\begin{itemize}
\item \verb+mf:STBoundedBy+ specifies the spatiotemporal bounding box and time units
\item \verb+mf:Member+  specifies properties of the moving features
\item \verb+mf:Header+  contains meta-information about the moving features, for example, attribute definition
\item \verb+mf:Foliation+ specifies the moving geometry and the dynamic attributes
\end{itemize}

While the XML encoding does support different interpolation types, the same interpolation applies to all attributes (linear by default). Therefore, we cannot specify that the geometry evolves linearly while the gear changes in a stepwise manner.

\begin{figure*}[htb]
\begin{lstlisting}
{
  "type": "MovingFeature",
  "temporalGeometry": {
      "type": "MovingPoint",
      "coordinates": [ [10.0, 10.0], [10.4, 11.2], [10.6, 12.2] ],
      "datetimes": ["2011-07-14T22:00:00Z", "2011-07-14T22:00:10Z", "2011-07-14T22:00:20Z"],
      "interpolations": "Linear"
  }, 
  "temporalProperties":  [ {
      "name": "gear",
      "values": [1, 2, 3, 3],
      "datetimes": ["2011-07-14T22:00:00Z", "2011-07-14T22:00:05Z", "2011-07-14T22:00:15Z",
                         "2011-07-14T22:00:20Z"],
      "interpolations": "Stepwise"
  }, ],
  "stBoundedBy": {
      "bbox": [10.0, 10.0, 10.6, 12.2],
      "period": { "begin": "2011-07-14T22:00:00Z", "end" : "2011-07-15T22:00:20Z" }
  }, 
  "properties": { 
      "name": "NissanA", "description": "Nissan Sentra ..." 
  }
}
\end{lstlisting}
\caption{JSON encoding of vehicle $A$ }
\label{fig:json}
\end{figure*}

\subsection{JSON Encoding}

Contrary to the previous two encodings, the JSON encoding is point based rather than segment based. Fig.~\ref{fig:json} shows the JSON representation of the example introduced in Fig.~\ref{fig:example}. The essential elements are:

\begin{itemize}
\item \verb+temporalGeometry+ represents the coordinates and timestamps in parallel arrays and states that linear interpolation is used
\item \verb+temporalProperties+ allows us to represent several such properties, where the values and the timestamps of each of them are represented as parallel arrays, and the interpolation is also stated, which is stepwise in our example for the attribute gear
\item \verb+stBoundedBy+ states the spatio-temporal bounding box
\item \verb+properties+ represents static / non-temporal descriptive attributes
\end{itemize}

The JSON encoding is considerably more compact than the CSV and XML encodings since the temporal geometry and the temporal attributes and their respective timestamps are encoded independently. In a real-world vehicle tracking use case, for example, there would considerably fewer gear changes than location changes. Therefore, it saves space to only record gear change events when necessary instead of reporting the gear setting at every location.

However, there are still ambiguous situations: for example, for attributes with stepwise interpolation, it is not possible to specify whether the change of value is exactly at the timestamp or just after it. Therefore, when trying to determine the period of time when the first gear is used, we cannot specify whether it was used during [2011-07-14 22:00:00, 2011-07-14 22:00:05] or [2011-07-14 22:00:00, 2011-07-14 22:00:05) (notice the right-open end of the period).
This has implications when computing topological predicates for moving geometries, because -- depending on the predicate considered -- the fact that a moving geometry is inside or crosses a boundary needs to take into account whether the bounds of the period are left- and right-inclusive or not.

\subsection{Moving Feature Encodings Summary}

As the examples in the previous sections show, the three Moving Feature encoding schemes implement different concepts for modeling movement data. This makes it hard to understand and use the standard since each encoding choice has different consequences. 

To summarize the situation, Table~\ref{tab:1} lists the advantages and limitations of the three Moving Feature encoding options. 
Overall, the JSON encoding (which is the most recent addition to the standard) is the most compact and has the fewest limitations.
However, it is worth mentioning that the segment based CSV and XML encodings (unlike the point based JSON encoding) allow us to represent temporal gaps due to the fact that every segment has both a start and an end timestamps. This is important from a modeling perspective since it makes it possible to represent, for example, that the GPS signal was lost while a moving car was inside a tunnel.

\subsection{Moving Features Access}

Besides encodings, Moving Features also standardizes functions for working with movement data. The corresponding standard is called OGC Moving Features Access \cite{OGC16-120r3} and it covers a wide range of functions for the retrieval of trajectory feature attributes, as well as operations between trajectories and geometry objects. Functions to retrieve feature attributes include basic functions to access the location, speed, or acceleration at a given time, or the subtrajectory between two timestamps. Conversely, there is also a function to extract the time at a given point. These and similar functions can be found in many existing tools, such as the ones analyzed by \cite{graser_movingpandas_2019}. 

Beyond these basic functions, however, the standard also defines more obscure functions, such as, for example, timeToDistance which “shall return a graph of the time to distance function as a set of curves in the Euclidean space consisting of coordinate pairs of time and distance”. The standard remains unclear as to how exactly this curved graph should be implemented.

\subsection{Discussion}

The limited success of OGC Moving Features so far may be due to its failure to involve a wide user base during its development stage. The lack of wide-spread engagement certainly limits public awareness of the standard's existence. Some of the limitations (particularly of the CSV and XML encodings) as well as ambiguities in the standard may be due to limited diversity of perspectives represented during the development of the standard. 

The decision to define multiple encoding standards that implement different concepts for modelling movement data further complicates the situation and potentially hinders the standard's adoption.

Another limiting factor is that, currently, there is no official reference implementation of OGC Moving Features. However, MobilityDB \cite{zimanyi2019mobilitydb, MobilityDB} builds upon the OGC Moving Features specification to add support for spatiotemporal objects to PostGIS databases. This effort may be considered a reality check to gauge the real-world applicability of the standard.

\section{Towards an API for Mobility Data Science}

Multiple notable recent publications, such as \cite{dodge_analysis_2016, demsar_analysis_2015}, have emphasized the need for a movement data science that bridges the boundaries of established domains, such as ecology, human mobility analytics for health and planning, or transportation research. Why then has there been such limited progress towards a common understanding and implementations of the key elements of an API for movement data analysis or even just a common data exchange format? The following conjectures summarize some of the main challenges that we have identified:

\textbf{Conjecture 1: The curse of movement data heterogeneity.} From high-resolution continuous tracking data of cooperative moving objects to sparse checkpoint-based trajectories with large spatiotemporal uncertainties, the flavours of movement datasets vary vastly. Furthermore, there is no consistent terminology \cite{graser_movingpandas_2019}: for example, terms such as trajectory, track, path, move, trip or travel may used to describe the same or different concepts related to movement. 
Moreover, while some domains still struggle to collect sufficient data, others are amassing huge amounts thanks to recent advances in location and communication technologies. But even if data is plentiful, there is a lack of commonly available data management tools for such volumes of data, and a pressing need for them. However, this heterogeneity means that it is unlikely that there will be one solution that fits all requirements. Therefore, our chances to build a flourishing mobility data science environment would be vastly improved by a shared terminology, as well as a set of standard data formats and well-defined analysis functions. 

\textbf{Conjecture 2: Lack of pragmatic solutions.} Many recently popularized data formats followed a somewhat opposite approach to established standardization processes. For example, GeoJSON, Vector tiles, or GTFS (General Transit Feed Specification) and the software tools around them quickly gained popularity as useful and pragmatic solutions for common problems. Individual mobility data projects, such as, for example, \cite{TraceJSON} try to follow a similar approach but have yet to reach a critical mass of users. The challenge lies in addressing the common problems faced by movement data analysts in different domains.  

\textbf{Conjecture 3: Wrong priorities and incentives.} Spatiotemporal data analysis is a hard nut to crack. However, there is no shortage of publications and sophisticated concepts, reaching back to, works such as \cite{sistla1997modeling}. Similarly, there are numerous research software prototypes but most of them are not usable since they were either never made publicly available or not engineered for use by others other than the original author. The current scientific environment provides very limited incentives for the development of well-engineered scientific software. It is risky to spend time on development when academic hiring committees may only look at the conventional publication record.

\section{Conclusion}

To solve the above mentioned issues and advance the mobility data science domain, we have to take steps to overcome these challenges. Looking back on how static/non-moving spatial data science tools have evolved over the last decades, we advocate for the following essential building blocks: 

\textbf{A. Common movement data analysis concepts:} Whether through official standards or by adopting a pragmatic de facto standard, a common terminology would vastly improve the exchange of movement data methodology. However, so far, there are no strong contenders for a widely adopted standard. OGC's Moving Feature Standards lacks essential features that would facilitate its wide-spread adoption. A well-accepted standard would provide a common reference framework for development efforts from all sides of the mobility data science community. Furthermore, it has the potential to facilitate the exchange of movement analysis methods by establishing a shared language with a well defined terminology. 

\textbf{B. An open general-purpose mobility (data) engine:} There are standard libraries for manipulating non-moving geometries (such as the JTS Topology Suite (Java Topology Suite) and its C++ port GEOS \cite{geos}) and numerous spatial data analysis tools (such as GeoTools, GDAL, R, PostGIS, QGIS, Google Earth, and NASA WorldWind) build on these libraries. Similarly, mobility data science would profit from a standard library for manipulating moving geometries. To avoid unnecessary duplication of spatial data handling functionality, such a mobility engine should build upon existing libraries for non-moving geometries. The mobility engine development should focus on the specific spatiotemporal aspects of movement data, while delegating the geometric manipulation to the underlying libraries. Language bindings for higher-level programming languages such as Python and R should be considered essential for a wider industry adoption. This general purpose mobility engine would enable faster development of more specialized movement data analysis tools while still keeping a common base framework. 

\textbf{C. Focus on open science and reproducibility:} A push towards open scientific practices, including code sharing and replication of results by peers (before and after publication) would be one step towards incentivizing the development of better scientific software. Furthermore, it would speed up research and development by removing the need to re-implement the same things over and over again and instead build on each others work.



\bibliographystyle{apalike}  
\bibliography{refs}

\end{document}